\documentclass[prl,twocolumn,showpacs,superscriptaddress]{revtex4}

\usepackage{amsmath}
\usepackage{verbatim}
\usepackage{graphicx}

\DeclareFontFamily{OT1}{rsfs}{}
\DeclareFontShape{OT1}{rsfs}{m}{n}{ <-7> rsfs5 <7-10> rsfs7 <10->rsfs10}{} 
\DeclareMathAlphabet{\mycal}{OT1}{rsfs}{m}{n}
\newcommand{\scri}{{\mycal I}}

\newcommand{\e}{\epsilon}

\renewcommand{\L}{{\mathcal{L}}}
\newcommand{\bL}{\bar{{\mathcal{L}}}}

\newcommand{\be}[1]{ \begin{equation}\label{#1} }
\newcommand{\ee}{\end{equation}}
\newcommand{\bea}[1]{\begin{eqnarray}\label{#1} }
\newcommand{\eea}{\end{eqnarray}}
\newcommand{\p}{\partial}
\newcommand{\D}{\Delta}

\newcommand{\refb}[1]{(\ref{#1})}

\renewcommand{\>}{\rangle}

\newcommand{\eq}[2]{\begin{equation} #1 \label{#2} \end{equation}}

\newcommand{\al}{\alpha}
\newcommand{\ga}{\gamma}
\newcommand{\de}{\delta}

\newcommand{\De}{\Delta}

\DeclareMathOperator{\extdm}{d}
\newcommand{\extd}{\extdm \!}

\begin{document}
 
\title{Flat-Space Chiral Gravity}

\author{Arjun Bagchi}
\email{arjun.bagchi@ed.ac.uk}
\affiliation{School of Mathematics, University of Edinburgh, Kings Buildings, Edinburgh EH9 3JZ, United Kingdom}

\author{St\'ephane Detournay}
\email{sdetourn@physics.harvard.edu}
\affiliation{Center for the Fundamental Laws of Nature, Harvard University, Cambridge, MA 02138, USA}

\author{Daniel Grumiller}
\email{grumil@hep.itp.tuwien.ac.at}
\affiliation{Institute for Theoretical Physics, Vienna University of Technology, Wiedner Hauptstrasse 8--10/136, A-1040 Vienna, Austria}

\date{\today}


\begin{abstract} 
We provide the first evidence for a holographic correspondence between a gravitational theory in flat space and a specific unitary field theory in one dimension lower.
The gravitational theory is a flat-space limit of topologically massive gravity in three dimensions at Chern--Simons level $k=1$.
The field theory is a chiral two-dimensional conformal field theory with central charge $c=24$.
\end{abstract}


\maketitle

One of the main pillars of our contemporary understanding of quantum gravity is the holographic principle \cite{'tHooft:1993gx}.
It states that a quantum theory of gravity in $d+1$ dimensions should have an equivalent description in terms of an ordinary unitary quantum (field) theory without gravity in $d$ dimensions.
The holographic principle had no concrete realization until Maldacena's seminal work on the Anti-deSitter/Conformal Field Theory (AdS/CFT) correspondence \cite{Maldacena:1997re}, which established that string theory in AdS incorporates the holographic principle in a specific way.
In order to understand holography better and for obvious practical purposes, one would like to formulate AdS/CFT-like scenarios in asymptotically flat spacetimes. 
This is the main purpose of our letter.

Even though progress has been achieved on various fronts in order to extract features of the flat space S-matrix from AdS/CFT correlators, see e.g.~\cite{Polchinski:1999ry}, it is fair to say that efforts at flat-space holography have not met with a great deal of success.
This is somewhat surprising, given that flat spaces can be obtained as a large radius limit of AdS~\footnote{%
\label{fn:1}
For example, starting with global AdS$_3$,
$\extd s^2 = \ell^2\,\big(\extd\rho^2-\cosh^2{\!\!\rho}\, \extd\tau^2 +\sinh^2{\!\!\rho}\,\extd\phi^2\big)$, 
defining  $u=\ell(\tau-\rho)$, $r=\ell\rho$, $\theta=\phi$ and taking the limit $\ell\to\infty$ (keeping finite $r$ and $u$) leads to the flat line-element \eqref{Mink} in the main text.}.
One could expect that holography for flat spaces should arise as a similar limit of the usual holographic dictionary in AdS. 

An important precursor of the AdS/CFT correspondence is the observation by
Brown and Henneaux \cite{Brown:1986nw} that any consistent theory of
three-dimensional (3D) quantum gravity with asymptotically AdS boundary
conditions is a 2D CFT, in the sense that 
the asymptotic symmetry group --- generated by all (non-trivial) diffeomorphisms that preserve
the asymptotic AdS boundary conditions ---  
is the 2D conformal group (see also \cite{Strominger:1997eq}). 
In flat space-times, the
asymptotic symmetry group is the infinite dimensional
Bondi--Metzner--Sachs (BMS) group \cite{Bondi:1962}. It is therefore natural to expect
the BMS group and its associated asymptotic symmetry algebra to play crucial roles in flat space
holography.

In pure 3D Einstein gravity the asymptotic symmetry algebra is the BMS$_3$ algebra \cite{ABS} and
picks up central extensions \cite{Barnich:2006av}. It is related to the conformal algebra through a redefinition of generators and taking the cosmological constant to zero \cite{Barnich:2006av, Bagchi:2010zz, Bagchi:2012cy}. 
Similarly, the general asymptotically flat solution to 3D Einstein gravity emerges as limit of the general solution in AdS \cite{Barnich:2010eb, Barnich:2012aw}. From the field theory point of view the limit of small cosmological constant is perceived as a contraction on the dual CFT  \cite{Bagchi:2010zz, Bagchi:2012cy}. 
So, if we believe that quantum gravity on AdS is dual to a CFT, the structure of the field theory dual for flat-space would be given by a contraction of a CFT.  Interestingly, these contracted CFTs were studied earlier in the context of non-relativistic limits of CFTs and are called Galilean conformal algebras (GCA) \cite{Bagchi:2009my}. This intriguing connection was dubbed the BMS/GCA correspondence \cite{Bagchi:2010zz}. 

So far the best understood example of this connection is in three bulk dimensions.
There the centrally extended BMS (or GCA) algebra is generated by Virasoro generators $L_n$ and supertranslations $M_n$ (with integer $n$).
\begin{subequations}
\label{eq:GCA} 
\begin{align}
[L_m,\,L_n]&= (m-n)\,L_{n+m} + \frac{c_1}{12}\,(n^3-n)\,\de_{n+m,0}  \label{eq:virasoro} \\
[L_m,\,M_n]&= (m-n)\,M_{n+m} + \frac{c_2}{12}\,(n^3-n)\,\de_{n+m,0} \\
[M_m,\,M_n]&= 0
\end{align}
\end{subequations}
In \cite{Bagchi:2012cy}, a precise spacetime picture for the limiting procedure was outlined which generated the flat space asymptotic symmetry algebra \eqref{eq:GCA} from the AdS asymptotic symmetry algebra. 

Even though 3D Einstein gravity is the simplest setup to address flat space holography, it comes with the major problem that there is no concrete proposal yet for a specific field theory with \eqref{eq:GCA} as asymptotic symmetry algebra and central extensions $c_1=0$, $c_2\neq 0$, as predicted from Einstein gravity \cite{Barnich:2006av}. The situation would be significantly better if $c_1\neq 0$ and $c_2=0$, since then the non-trivial part of the asymptotic symmetry algebra \eqref{eq:GCA} would reduce to one copy of the Virasoro algebra and one may expect (the chiral half of) a standard CFT as field theory dual.

In this letter we solve this problem.  Namely, we show that a possible way around is to add to pure Einstein gravity a gravitational Chern-Simons term. 
The theory is called Topologically Massive Gravity (TMG) \cite{Deser:1982vy}. 
The action of TMG is given by 
\be{TMG}
I_{\textrm{\tiny TMG}} = \frac{1}{16\pi G}\,\int \extd^3x \sqrt{-g}\,\Big[  R + \frac{2}{\ell^2} + \frac{1}{2\mu}\, \textrm{CS}(\Gamma) \Big]
\ee
where $G$ is the Newton constant, 
$R$ the Ricci scalar, $\ell$ the AdS radius, $\mu$ is the Chern--Simons coupling and CS$(\Gamma)=\varepsilon^{\lambda \mu \nu} \Gamma^\rho{}_{\lambda \sigma} \big( \partial_\mu \Gamma^\sigma{}_{\rho\nu} + \frac{2}{3} \Gamma^\sigma{}_{\mu\tau} \Gamma^{\tau}{}_{\nu \rho} \big)$ is the gravitational Chern--Simons term.

The asymptotic symmetries of 3D flat space at null infinity were studied for Einstein gravity in \cite{Barnich:2006av}. 
We impose boundary conditions on the metric $g_{\mu\nu}$, which generalize/correct the ones proposed in \cite{Barnich:2006av}: 
\begin{subequations}
 \label{BMSBC}
\begin{align}
 g_{uu} &= h_{uu} + O(\tfrac 1r) \qquad g_{ur} = -1 + h_{ur}/r + O(\tfrac{1}{r^2}) \\ 
 g_{u\theta} &= h_{u\theta} + O(\tfrac 1r) \qquad g_{rr} = h_{rr}/r^2+O(\tfrac{1}{r^3}) \\ 
 g_{r\theta} &= h_1 (\theta)  + h_{r\theta}/r + O(\tfrac{1}{r^2}) \\
 g_{\theta\theta} &= r^2 + (h_2(\theta) + u h_3(\theta)) r + O(1)
\end{align}
\end{subequations}
All coefficients $h_{\mu\nu}$ are functions of retarded time $u$ and angle $\theta$, but they do not depend on the radius $r$.
For $h_{uu}=-1$ we recover asymptotically the Minkowski line-element in outgoing Eddington--Finkelstein coordinates:
\be{Mink}
\extd s^2 = -\extd u^2 - 2 \extd r \extd u + r^2 \extd\theta^2
\ee 
In these coordinates the future null boundary $\scri^+$ is approached in the limit $r\to \infty$. 
The six Killing vectors of \eqref{Mink}, $\ell_n$, $m_n$ (with $n=\pm 1,0$), form an $iso(2,1)$ algebra: 
\begin{subequations}
\label{eq:killings}
 \begin{align}
  \ell_n &= i e^{i n \theta} (in u \p_u - in (r+u)\p_r + (1+ n^2 \frac{u}{r})\p_\theta)\\
   m_n &= i e^{i n \theta} (\p_u - n^2 \p_r - i \frac{n}{r}\p_\theta)
 \end{align}
\end{subequations}
The asymptotic symmetry group is generated by
\begin{subequations}
 \label{eq:prl1}
\begin{align}
 L_n &= i e^{i n \theta} \big(i n u \p_u - i n r\p_r +(1+ n^2 \frac{u}{r}) \p_\theta) + \dots\\ 
 M_n &= i e^{i n \theta} \p_u +\dots
\end{align}
\end{subequations}
where the dots refer to sub-leading terms, $ \dots = {\cal O}(\tfrac1r) \p_u + \big(u f_1(\theta) + f_2(\theta) + {\cal O}(\tfrac 1r)\big)\p_r + \big(f_3(\theta)/r + {\cal O}(\tfrac{1}{r^2})\big) \p_\theta$, generating trivial gauge transformations which are modded out in the asymptotic symmetry group.
The generators preserve the boundary conditions \eqref{BMSBC} and satisfy asymptotically the BMS algebra \eqref{eq:GCA} (without central terms). The generators $L_{\pm 1}$ , $M_{\pm 1}$ and the corresponding Killing vectors $\ell_{\pm 1}$, $m_{\pm 1}$ differ by a trivial gauge transformation.

The boundary conditions \eqref{BMSBC} can be verified (e.g. using \cite{CodeGeo}) to be consistent in TMG, yielding well-defined charges that are finite, integrable and conserved. 
They are given by 
\begin{subequations}
\label{ChargesBMSTMG}
 \begin{align}
  Q_{M_n} &= \frac{1}{16 \pi G}\, \int \extd\theta\, e^{i n \theta}
 \; \big(h_{uu} + h_3\big) \,,  \\
  Q_{L_n} &= \frac{1}{16 \pi G \mu}\, \int\extd\theta\, e^{i n \theta} \; \big(h_{uu} + \p_u h_{ur} + \tfrac{1}{2} \p_u^2 h_{rr} + h_3\big)
\nonumber\\
& \!\!\!\!+ \frac{1}{16 \pi G}\, \int \extd\theta\, e^{i n \theta}
 \; \big( i n u h_{uu} + i n h_{ur} + 2 h_{u\theta} + \p_u h_{r\theta} \nonumber \\ 
 & \qquad\qquad\quad - (n^2+ h_3 ) h_1 -  i n h_2 - i n \p_\theta h_1      \big)          \,.
\end{align}
\end{subequations}
The proof of the conservation of the charges requires to solve the equations of motion (EOM) asymptotically.
For finite values of $\mu$ the crucial on-shell conditions that establish charge conservation are given by
$\partial_u h_{uu} = 0$, $h_{ur} = -\frac12\,\partial_u h_{rr}$, and $u\partial_\theta h_{uu} + \partial_\theta h_{ur} = 2h_{u\theta} + \partial_u h_{r\theta}$.

When realized as asymptotically conserved charges, the charge algebra picks up central extensions exactly as in \eqref{eq:GCA}.
In TMG we obtain
\be{cc}
c_1 = \frac{3}{\mu G}\,, \qquad c_2 = \frac{3}{G}\,.
\ee
This is consistent with the result of \cite{Barnich:2006av} for Einstein gravity, recovered in the limit $\mu \to \infty$. 
The point of interest here is that the Virasoro part of the BMS algebra acquires a non-trivial central extension, $c_1\neq 0$.
This resolves one of the problems encountered in Einstein gravity. 

In order to resolve another one, namely the fact that $c_2\neq 0$, we briefly reconsider TMG in AdS.
This is also useful in its own right, since the BMS algebra \eqref{eq:GCA} can be obtained by a contraction of the AdS asymptotic symmetry algebra, 
$L_n = \L_n - \bL_{-n}$, $M_n =\frac{1}{\ell}\, (\L_n + \bL_{-n})$ \cite{Barnich:2006av}.
Here $\ell$ is the AdS-radius, which is sent to infinity in the flat-space limit.
The $\L_n$ and $\bL_n$ are the generators of two copies of the Virasoro algebra [see \eqref{eq:virasoro}], with central charges
$c = \frac{3 \ell}{2G}\, ( 1 + \frac{1}{\mu \ell})$, $\bar c = \frac{3 \ell}{2G}\, ( 1 - \frac{1}{\mu \ell})$ \cite{Kraus:2005zm}.
In the limit $\ell \to \infty$ the corresponding BMS central charges in the algebra \eqref{eq:GCA} then become
$c_1 = \lim_{\ell \to \infty} ( c - \bar c) = \frac{3}{\mu G}$, $c_2 = \lim_{\ell \to \infty} \frac{1}{\ell}( c + \bar c) =  \frac{3}{G}$.
This agrees precisely with the result \eqref{cc} of our canonical analysis.

The consistency check we just performed indicates how to proceed to obtain vanishing central charge $c_2$:
We should consider a limit of TMG where $c = -\bar c$.
Alternatively, one can take the flat-space limit of TMG where additionally Newton's constant is scaled to infinity, $G\to\infty$, while keeping fixed $\mu G$:
\be{eq:42}
\mu=\e\to 0\,, \quad G = \frac{1}{8k \e}\to\infty \quad \mbox{so that} \quad \mu G = \frac{1}{8k} 
\ee 
The quantity $k$ is the rescaled inverse Newton constant, whose meaning will become clear in a moment.
Both limits described above exist and both lead to conformal Chern--Simons gravity (CSG) \cite{Afshar} 
with action
\be{CSG}
I_{\textrm{\tiny CSG}} = \frac{k}{4\pi}\int \extd^3x \sqrt{-g}\, \textrm{CS}(\Gamma)\,.
\ee
The constant $k$ is now recognized as the Chern--Simons level.

It is known that CSG \eqref{CSG} admits flat solutions \cite{Deser:1982vy}. 
So the flat limit of TMG \eqref{TMG} in the scaling limit \eqref{eq:42} is CSG \eqref{CSG}. 
The dual of this theory, if it exists, is given by the 2D GCA \eqref{eq:GCA} with central charges  [see \eqref{cc}]
\eq{
c_1 = 24 k \qquad c_2 = 0\,.
}{eq:centralcharges}
Both central charges are of the desired form. 
This is one of the main results of this letter.

It is worthwhile to compare with the situation of AdS holography in TMG.
There the only candidate for a unitary theory (with Brown--Henneaux boundary conditions) and macroscopic central charge ($c>1$) is chiral gravity \cite{Li:2008dq}.
Interestingly, chiral gravity also leads to a single copy of the Virasoro algebra as asymptotic symmetry algebra, which suggests that chiral gravity could be related to our flat-space limit of TMG with the scaling limit \eqref{eq:42} [or, equivalently, of CSG \eqref{CSG} with our flat-space boundary conditions \eqref{BMSBC}].
We now strengthen the analogy with chiral gravity arguing that our resulting theory is indeed chiral.

As demonstrated above the states of the bulk form representations of the 2D GCA. Such representations are labeled by the eigenvalues $\xi$ and $\De$ of $L_0$ and $M_0$ 
\cite{Bagchi:2009ca,Bagchi:2009pe},
 $L_0 | \Delta, \xi \rangle = \xi | \Delta, \xi \rangle$,  $M_0 | \Delta, \xi \rangle = \Delta | \Delta, \xi \rangle$. 
One defines the notion of primary states in the usual CFT language as the states annihilated by $L_n, M_n$ for $n>0$. 
The representations are built by acting on these primary states with raising operators $L_{-n}, M_{-n}$, 
which
raises the $\xi$ eigenvalue 
to $\xi + n$. 
For CSG \eqref{CSG} the eigenvalue $\Delta$ vanishes in the flat-space limit. 
From general considerations 
\cite{Bagchi:2009pe} we know that the 2D GCA \eqref{eq:GCA} with $c_2=0$ has unitary sub-sectors for $\Delta=0$ where the GCA module can be reduced to the Virasoro module and usual unitary requirements of 2D CFTs apply here. 
All this fits nicely with the suggestion that the dual of chiral gravity is the chiral half of a CFT. 

Therefore, we call CSG \eqref{CSG} with our boundary conditions \eqref{BMSBC} ``flat-space chiral gravity'' and conjecture that it is dual to a chiral half of a CFT with central charge $c=24k$.
We discuss now several important consequences and additional checks of our conjecture.

We mentioned that for $c_2=\Delta=0$ the representations of the 2D GCA reduce to those of the Virasoro algebra. 
By analyzing null vectors following \cite{Bagchi:2009pe} we substantiate now this claim. 
Like in usual 2D CFTs, null states in the GCA representations are states which are orthogonal to all states including themselves. 
At level one the most general state is given by
$(a L_{-1} + b M_{-1}) | \xi, \D \>$.
Acting with $L_1$ or $M_1$ and requiring the results to vanish gives the conditions
$\D = a = 0$ (or the trivial $a=b=0$). 
The level one null state is then given by 
$| \chi_1 \> = M_{-1} | \xi, 0 \>$.
At level two the most general state is 
$\big( a_1 L_{-2} + a_2 L_{-1}^2 + b_1 L_{-1}M_{-1} + d_1 M_{-1}^2 + d_2 M_{-2} \big) | \xi, \D \>$. 
Again we find the conditions for the existence of null states  
by acting with lowering operators. 
We restrict our attention to the case 
$c_2=\Delta=0$.
Then the constraints for the existence of null vectors simplify to
$a_1 = a_2 =0$ and $b_1 = - \tfrac{3}{2(\xi+1)} d_2$.
This leads to two level two null vectors,
$| \chi_2^1 \> = M_{-1}^2   | \xi, 0 \>$ and  $| \chi_2^2 \> = ( L_{-1}M_{-1}  - \tfrac23 (\xi+1)M_{-2}) | \xi, 0 \>$.
There are no constraints on 
the central charge $c_1$ or the weight $\xi$. 
If we consistently set the level one null state $| \chi_1 \>$ and its descendants to zero, at level two we are left with just 
$| \chi_2 \> = M_{-2} | \xi, 0 \>$.

This analysis can be continued, and at any level $n$ we find that we have a new null state $| \chi_n \> = M_{-n} | \xi, 0 \>$ if we set all the other lower level null states and their descendants to zero. 
Thus, if all the null states are truncated in a consistent manner, the tower of states precisely reduces to the Virasoro tower given by the Virasoro descendants of the primary state. 
There is generically no condition on $c_1$ and $\xi$; hence we can consider the truncation of the Hilbert space to just the Virasoro module. Here by the usual analysis of null vectors of the Virasoro algebra, one can put unitarity constraints on the values of central charge $c_1$ and weight $\xi$. 
In conclusion, we can have unitary representations of the GCA with $c_2=\De=0$. 
We call this the chiral truncation of the GCA. 

We study next aspects of the linearized spectrum by considering solutions $\psi_{\mu\nu}$ to the linearized CSG EOM around the flat background \eqref{Mink}.
A class of such solutions parametrized by the $L_0$-eigenvalue $\xi$ is given by 
\eq{
\psi_{\mu\nu}(\xi) = 
e^{-i(\xi+2)\theta}r^{-\xi-2}\, \big(m_1\otimes m_1\big)_{\mu\nu}
}{eq:psi}
The modes \eqref{eq:psi} are primaries, in the sense that $\ell_1 \psi=0=m_1\psi$, traceless, since $(m_1)^2=0$, and transverse, $\nabla_\mu\psi^{\mu\nu}=0$. They obey the differential equation
${\cal D}\psi_{\al\beta} := \epsilon_\al{}^{\ga\de}\nabla_\ga \psi_{\de\beta} = 0$ and are a flat-space analogue of the AdS modes constructed in \cite{Li:2008dq}.
In transverse gauge the linearized CSG EOM reduce to $({\cal D}\psi)^3 = 0$.
There are two additional branches of solutions \cite{Carlip:2008jk}.
One is the flat-space analogue of log excitations \cite{Grumiller:2008qz},
$\psi^{\rm log}_{\al\beta}(\xi) = -i(u+r)\,\psi_{\al\beta}(\xi)$.
The other one could be called log-squared,
 $\psi^{\rm log2}_{\al\beta}(\xi) = -{\tfrac{1}{2}}(u+r)^2\,\psi(\xi)$.
The angular momentum operator $L_0$ expectedly is diagonal. 
The operator $M_0$ has a rank-3 Jordan cell:
\eq{
M_0 \,\left(\begin{array}{c} \psi^{\rm log2} \\ \psi^{\rm log} \\ \psi \end{array}\right) =
\left(\begin{array}{ccc} 
0 & 1 & 0 \\
0 & 0 & 1 \\
0 & 0 & 0
\end{array}\right) \,\left(\begin{array}{c} \psi^{\rm log2} \\ \psi^{\rm log} \\ \psi \end{array}\right) 
}{eq:m0}
The result \eqref{eq:m0} differs from AdS-TMG \cite{Grumiller:2008qz} and flat-space Einstein gravity, which have rank-2 Jordan cells.

An even more crucial difference to the AdS case is the fact that all modes \eqref{eq:psi} (and their log and log-squared partners, as well as their $\ell_{-1}$ and $m_{-1}$ descendants) are either divergent at $r=0$ or incompatible with our boundary conditions \eqref{BMSBC}. Moreover, the modes compatible with our boundary conditions all have vanishing charges. However, we can construct directly linearized modes by acting with the Virasoro generators $L_{-n}$ on the vacuum \eqref{Mink}, denoted by $\psi^{(n)}=L_{-n}|0\>$ on the CFT side.
We obtain on the gravity side the non-vanishing components
\eq{
\psi^{(n)}_{uu} = -2n e^{-in\theta},\;
\psi^{(n)}_{u\theta} = -\frac{inu}{2}\psi^{(n)}_{uu},\; 
\psi^{(n)}_{\theta\theta} = -n^2 ru \psi^{(n)}_{uu}.
}{eq:primary}
Note that the modes \eqref{eq:primary} are neither traceless nor transverse, but solve the linearized CSG EOM, are compatible with our boundary conditions and regular in the interior.
The conserved Virasoro charges are given by $Q_{L_n}(\psi^{(n)})=2k(n^3-n)$; in particular, $Q_{L_n}(\psi^{(0,\pm 1)})=0$ as expected from the CFT side. Thus, the spectrum of the modes \eqref{eq:primary} consistently matches the spectrum of Virasoro descendants of the vacuum on the CFT side.

\newcommand{\parameter}{\alpha} 

In addition to the linearized spectrum it is of interest to look for non-perturbative states, basically the flat-space limit of BTZ black holes \cite{Banados:1992wn, BDRS, BarnichToapp}. They are important contributions to the quantum gravity partition function and required for modular invariance \cite{Maloney:2007ud}. We find indeed such solutions compatible with 
\eqref{BMSBC}, parametrized by the locus of the Killing horizon $r_0$ and a parameter $\parameter$:
\eq{
\extd s^2 = \parameter^2\,\big(1-\tfrac{r_0^2}{r^2}\big)\,\extd u^2 - 2\extd u\extd r + r^2 \,\big(\extd\theta - \tfrac{\parameter r_0}{r^2}\,\extd u\big)^2
}{eq:BTZ}

We calculate now the conserved charges associated with these ``flat BTZ'' solutions \eqref{eq:BTZ}.
In the limit \eqref{eq:42}, the charges of the asymptotic symmetry algebra \refb{ChargesBMSTMG} simplify and the previous on-shell conditions are replaced by $\partial_u h_{uu} + \partial_u^2 h_{ur} + \frac12\,\partial_u^3 h_{rr} = 0$.
This is sufficient to prove conservation of the charges $Q_{L_n}$, which read explicitly 
\eq{
 Q_{L_n} =  \frac{k}{2 \pi}\, \int\extd\theta\, e^{i n \theta} \; \big(h_{uu} + \p_u h_{ur}+ \tfrac{1}{2} \p_u^2 h_{rr} + h_3\big)\,. 
}{eq:QLn}
The charges $Q_{M_n}$ vanish due to the scaling limit \eqref{eq:42}.
From \eqref{eq:QLn} we read off that the vacuum \eqref{Mink} has the conserved Virasoro charge
$Q_{L_0} (\textrm{vacuum})= -k$
while the non-perturbative states \eqref{eq:BTZ} have
$Q_{L_0} (\textrm{flat\;BTZ})= k \parameter^2$.

Thus, we have just proven that there is a gap in the spectrum between the vacuum and the non-perturbative states. The size of the gap is given by $k=\tfrac{c}{24}$.
The presence of this gap is a non-trivial check of our conjecture, and an indication \cite{Witten:2007kt} that the dual CFT might be an extremal CFT.
In particular, for $k=1$ Witten has identified a specific extremal CFT with $c=24$  \cite{Witten:2007kt}, namely the monster theory of Frenkel, Lepowsky and Meurman \cite{Frenkel}. This allows us to sharpen our conjecture, namely that flat-space chiral gravity at Chern--Simons level $k=1$ is dual to the monster CFT.
If this is true and also the chiral gravity conjecture \cite{Li:2008dq} holds then 
chiral gravity and flat-space chiral gravity must be dual to each other. 
Alternatively, either of the purported gravity duals of the monster CFT could be unstable and decay/flow to the stable solution.
The RG analysis in Ref.~\cite{Percacci:2010yk} is an indication that our limit \eqref{eq:42} is stable under RG flow, while the chiral gravity tuning $\mu\ell=1$ is not.

It is of interest to perform further checks of our conjecture. We mention some promising avenues. Thermodynamics should be studied and consistency with the Cardy formula should be checked. The quantum gravity partition function should be calculated and compared with CFT expectations. Two- and three-point correlators can be calculated on the gravity side to check consistency with conformal Ward identities. 
Finally, it would be good to clarify if other flat-space limits are possible, for instance limits that do not localize asymptotically on the future null boundary, but that include also other components of the asymptotic boundary, or similar limits in other theories of gravity. We intend to address some of these aspects in the future \cite{BDG}.


\acknowledgments

We are grateful to H.~Afshar, T.~Andrade, G.~Barnich, A.~Castro, G.~Comp\`ere, S.~Deser, R.~Fareghbal, M.~Gary, G.~Giribet, R.~Gopakumar, T.~Hartman, D.~Hofman, R.~Jackiw, R.~Loganayagam, A.~Maloney, G.S.~Ng, I.~Sachs, J.~Simon and A.~Strominger for discussions.
We thank KITP for hospitality during the program ``Bits, Branes, Black Holes''. 
AB also thanks Harvard University and Brown University for hospitality during the course of this work. 
AB was supported by EPSRC. 
SD was supported by the Fundamental Laws Initiative of the Center for the Fundamental Laws of Nature, Harvard University. DG was supported by the START project Y435-N16 of the Austrian Science Fund (FWF) and the FWF project P21927-N16.
This research was supported in part by the National Science Foundation under Grant No. NSF PHY11-25915 and by an Excellence Fellowship of Wallonie-Bruxelles International.


\begin{thebibliography}{31}
\expandafter\ifx\csname natexlab\endcsname\relax\def\natexlab#1{#1}\fi
\expandafter\ifx\csname bibnamefont\endcsname\relax
  \def\bibnamefont#1{#1}\fi
\expandafter\ifx\csname bibfnamefont\endcsname\relax
  \def\bibfnamefont#1{#1}\fi
\expandafter\ifx\csname citenamefont\endcsname\relax
  \def\citenamefont#1{#1}\fi
\expandafter\ifx\csname url\endcsname\relax
  \def\url#1{\texttt{#1}}\fi
\expandafter\ifx\csname urlprefix\endcsname\relax\def\urlprefix{URL }\fi
\providecommand{\bibinfo}[2]{#2}
\providecommand{\eprint}[2][]{\url{#2}}

\bibitem[{\citenamefont{'t~Hooft}(1993)}]{'tHooft:1993gx}
\bibinfo{author}{\bibfnamefont{G.}~\bibnamefont{'t~Hooft}}, in
  \emph{\bibinfo{booktitle}{Salamfestschrift}} (\bibinfo{publisher}{World
  Scientific}, \bibinfo{year}{1993}), \eprint{gr-qc/9310026};
\bibinfo{author}{\bibfnamefont{L.}~\bibnamefont{Susskind}},
  \bibinfo{journal}{J. Math. Phys.} \textbf{\bibinfo{volume}{36}},
  \bibinfo{pages}{6377} (\bibinfo{year}{1995}),
  \eprint[http://arXiv.org/abs]{hep-th/9409089}.

\bibitem[{\citenamefont{Maldacena}(1998)}]{Maldacena:1997re}
\bibinfo{author}{\bibfnamefont{J.~M.} \bibnamefont{Maldacena}},
  \bibinfo{journal}{Adv. Theor. Math. Phys.} \textbf{\bibinfo{volume}{2}},
  \bibinfo{pages}{231} (\bibinfo{year}{1998}), \eprint{hep-th/9711200}.

\bibitem[{\citenamefont{Polchinski}(1999)}]{Polchinski:1999ry}
\bibinfo{author}{\bibfnamefont{J.}~\bibnamefont{Polchinski}}
  (\bibinfo{year}{1999}), \eprint{hep-th/9901076};
\bibinfo{author}{\bibfnamefont{L.}~\bibnamefont{Susskind}},
  \eprint{hep-th/9901079};
\bibinfo{author}{\bibfnamefont{S.~B.} \bibnamefont{Giddings}},
  \bibinfo{journal}{Phys.Rev.} \textbf{\bibinfo{volume}{D61}},
  \bibinfo{pages}{106008} (\bibinfo{year}{2000}), \eprint{hep-th/9907129};
\bibinfo{author}{\bibfnamefont{M.}~\bibnamefont{Gary}},
  \bibinfo{author}{\bibfnamefont{S.~B.} \bibnamefont{Giddings}},
  \bibnamefont{and}
  \bibinfo{author}{\bibfnamefont{J.}~\bibnamefont{Penedones}},
  \bibinfo{journal}{Phys.Rev.} \textbf{\bibinfo{volume}{D80}},
  \bibinfo{pages}{085005} (\bibinfo{year}{2009}), \eprint{0903.4437};
\bibinfo{author}{\bibfnamefont{M.}~\bibnamefont{Gary}} \bibnamefont{and}
  \bibinfo{author}{\bibfnamefont{S.~B.} \bibnamefont{Giddings}},
  \bibinfo{journal}{Phys.Rev.} \textbf{\bibinfo{volume}{D80}},
  \bibinfo{pages}{046008} (\bibinfo{year}{2009}), \eprint{0904.3544}.

\bibitem[{\citenamefont{Brown and Henneaux}(1986)}]{Brown:1986nw}
\bibinfo{author}{\bibfnamefont{J-D.}~\bibnamefont{Brown}} \bibnamefont{and}
  \bibinfo{author}{\bibfnamefont{M.}~\bibnamefont{Henneaux}}
\bibinfo{journal}{Commun.Math.Phys.} \textbf{\bibinfo{volume}{104}},
  \bibinfo{pages}{207-226} (\bibinfo{year}{1986}).


\bibitem[{\citenamefont{Strominger}(1997)}]{Strominger:1997eq}
\bibinfo{author}{\bibfnamefont{A. }~\bibnamefont{Strominger}}
\bibinfo{journal}{JHEP} \textbf{\bibinfo{volume}{9802}},
  \bibinfo{pages}{009} (\bibinfo{year}{1998}),\eprint{hep-th/9712251}.


\bibitem[{\citenamefont{Bondi et~al.}(1962)\citenamefont{Bondi, van~der Burg,
  and Metzner}}]{Bondi:1962}
\bibinfo{author}{\bibfnamefont{H.}~\bibnamefont{Bondi}},
  \bibinfo{author}{\bibfnamefont{M.}~\bibnamefont{van~der Burg}},
  \bibnamefont{and} \bibinfo{author}{\bibfnamefont{A.}~\bibnamefont{Metzner}},
  \bibinfo{journal}{Proc. Roy. Soc. London} \textbf{\bibinfo{volume}{A269}},
  \bibinfo{pages}{21} (\bibinfo{year}{1962});
\bibinfo{author}{\bibfnamefont{R.}~\bibnamefont{Sachs}},
  \bibinfo{journal}{Phys. Rev.} \textbf{\bibinfo{volume}{128}},
  \bibinfo{pages}{2851} (\bibinfo{year}{1962}).

\bibitem[{\citenamefont{A. Ashtekar et~al.}(1997)}]{ABS}
\bibinfo{author}{\bibfnamefont{A.}~\bibnamefont{Ashtekar}},
  \bibinfo{author}{\bibfnamefont{J.}~\bibnamefont{Bicak}}~\bibnamefont{and}
    \bibinfo{author}{\bibfnamefont{B. G.}~\bibnamefont{Schmidt}}  
  \bibinfo{journal}{Phys. Rev. D} \textbf{\bibinfo{volume}{55}},
  \bibinfo{pages}{669} (\bibinfo{year}{1997}), \eprint{gr-qc/9608042}.



\bibitem[{\citenamefont{Barnich and Compere}(2007)}]{Barnich:2006av}
\bibinfo{author}{\bibfnamefont{G.}~\bibnamefont{Barnich}} \bibnamefont{and}
  \bibinfo{author}{\bibfnamefont{G.}~\bibnamefont{Compere}},
  \bibinfo{journal}{Class.Quant.Grav.} \textbf{\bibinfo{volume}{24}},
  \bibinfo{pages}{F15} (\bibinfo{year}{2007}), \eprint{gr-qc/0610130}.

\bibitem[{\citenamefont{Bagchi}(2010)}]{Bagchi:2010zz}
\bibinfo{author}{\bibfnamefont{A.}~\bibnamefont{Bagchi}},
  \bibinfo{journal}{Phys.Rev.Lett.} \textbf{\bibinfo{volume}{105}},
  \bibinfo{pages}{171601} (\bibinfo{year}{2010}), \eprint{1006.3354}.


\bibitem[{\citenamefont{Bagchi and Fareghbal}(2012)}]{Bagchi:2012cy}
\bibinfo{author}{\bibfnamefont{A.}~\bibnamefont{Bagchi}} \bibnamefont{and}
  \bibinfo{author}{\bibfnamefont{R.}~\bibnamefont{Fareghbal}}
  (\bibinfo{year}{2012}), \eprint{1203.5795}.

\bibitem[{\citenamefont{Barnich and Troessaert}(2010)}]{Barnich:2010eb}
\bibinfo{author}{\bibfnamefont{G.}~\bibnamefont{Barnich}} \bibnamefont{and}
  \bibinfo{author}{\bibfnamefont{C.}~\bibnamefont{Troessaert}},
  \bibinfo{journal}{JHEP} \textbf{\bibinfo{volume}{1005}}, \bibinfo{pages}{062}
  (\bibinfo{year}{2010}), \eprint{1001.1541}; 



\bibitem[{\citenamefont{Barnich et~al.}(2012)\citenamefont{Barnich, Gomberoff,
  and Gonzalez}}]{Barnich:2012aw}
\bibinfo{author}{\bibfnamefont{G.}~\bibnamefont{Barnich}},
  \bibinfo{author}{\bibfnamefont{A.}~\bibnamefont{Gomberoff}},
  \bibnamefont{and} \bibinfo{author}{\bibfnamefont{H.~A.}
  \bibnamefont{Gonzalez}} (\bibinfo{year}{2012}), \eprint{1204.3288}.



\bibitem[{\citenamefont{Bagchi and Gopakumar}(2009)}]{Bagchi:2009my}
\bibinfo{author}{\bibfnamefont{A.}~\bibnamefont{Bagchi}} \bibnamefont{and}
  \bibinfo{author}{\bibfnamefont{R.}~\bibnamefont{Gopakumar}},
  \bibinfo{journal}{JHEP} \textbf{\bibinfo{volume}{0907}}, \bibinfo{pages}{037}
  (\bibinfo{year}{2009}), \eprint{0902.1385}.





\bibitem[{\citenamefont{Deser et~al.}(1982{\natexlab{a}})\citenamefont{Deser,
  Jackiw, and Templeton}}]{Deser:1982vy}
\bibinfo{author}{\bibfnamefont{S.}~\bibnamefont{Deser}},
  \bibinfo{author}{\bibfnamefont{R.}~\bibnamefont{Jackiw}}, \bibnamefont{and}
  \bibinfo{author}{\bibfnamefont{S.}~\bibnamefont{Templeton}},
  \bibinfo{journal}{Phys. Rev. Lett.} \textbf{\bibinfo{volume}{48}},
  \bibinfo{pages}{975} (\bibinfo{year}{1982}{\natexlab{a}});
  \bibinfo{journal}{Ann. Phys.} \textbf{\bibinfo{volume}{140}},
  \bibinfo{pages}{372} (\bibinfo{year}{1982}{\natexlab{b}}).


\bibitem[{\citenamefont{CodeGeo}()\citenamefont{Compere}}]{CodeGeo}
\bibinfo{author}{\bibfnamefont{G.}~\bibnamefont{Comp\`ere}},
  \bibinfo{note}{A Mathematica package for computing charges, http://staff.science.uva.nl/$\sim$compere/pack\-age.html}.


\bibitem[{\citenamefont{Kraus and Larsen}(2006)}]{Kraus:2005zm}
\bibinfo{author}{\bibfnamefont{P.}~\bibnamefont{Kraus}} \bibnamefont{and}
  \bibinfo{author}{\bibfnamefont{F.}~\bibnamefont{Larsen}},
  \bibinfo{journal}{JHEP} \textbf{\bibinfo{volume}{01}}, \bibinfo{pages}{022}
  (\bibinfo{year}{2006}), \eprint{hep-th/0508218}.

\bibitem[{\citenamefont{Afshar et~al.}(2011)\citenamefont{Afshar, Cvetkovic,
  Ertl, Grumiller, and Johansson}}]{Afshar}
\bibinfo{author}{\bibfnamefont{H.}~\bibnamefont{Afshar}},
  \bibinfo{author}{\bibfnamefont{B.}~\bibnamefont{Cvetkovic}},
  \bibinfo{author}{\bibfnamefont{S.}~\bibnamefont{Ertl}},
  \bibinfo{author}{\bibfnamefont{D.}~\bibnamefont{Grumiller}},
  \bibnamefont{and}
  \bibinfo{author}{\bibfnamefont{N.}~\bibnamefont{Johansson}},
  \bibinfo{journal}{Phys.Rev.} \textbf{\bibinfo{volume}{D84}},
  \bibinfo{pages}{041502(R)} (\bibinfo{year}{2011}), \eprint{1106.6299};
  \bibinfo{journal}{Phys.Rev.} \textbf{\bibinfo{volume}{D85}},
  \bibinfo{pages}{064033} (\bibinfo{year}{2012}), \eprint{1110.5644}.

\bibitem[{\citenamefont{Li et~al.}(2008)\citenamefont{Li, Song, and
  Strominger}}]{Li:2008dq}
\bibinfo{author}{\bibfnamefont{W.}~\bibnamefont{Li}},
  \bibinfo{author}{\bibfnamefont{W.}~\bibnamefont{Song}}, \bibnamefont{and}
  \bibinfo{author}{\bibfnamefont{A.}~\bibnamefont{Strominger}},
  \bibinfo{journal}{JHEP} \textbf{\bibinfo{volume}{04}}, \bibinfo{pages}{082}
  (\bibinfo{year}{2008}), \eprint{0801.4566}.

\bibitem[{\citenamefont{Bagchi and Mandal}(2009)}]{Bagchi:2009ca}
\bibinfo{author}{\bibfnamefont{A.}~\bibnamefont{Bagchi}} \bibnamefont{and}
  \bibinfo{author}{\bibfnamefont{I.}~\bibnamefont{Mandal}},
  \bibinfo{journal}{Phys.Lett.} \textbf{\bibinfo{volume}{B675}},
  \bibinfo{pages}{393} (\bibinfo{year}{2009}), \eprint{0903.4524}.

\bibitem[{\citenamefont{Bagchi et~al.}(2010)\citenamefont{Bagchi, Gopakumar,
  Mandal, and Miwa}}]{Bagchi:2009pe}
\bibinfo{author}{\bibfnamefont{A.}~\bibnamefont{Bagchi}},
  \bibinfo{author}{\bibfnamefont{R.}~\bibnamefont{Gopakumar}},
  \bibinfo{author}{\bibfnamefont{I.}~\bibnamefont{Mandal}}, \bibnamefont{and}
  \bibinfo{author}{\bibfnamefont{A.}~\bibnamefont{Miwa}},
  \bibinfo{journal}{JHEP} \textbf{\bibinfo{volume}{1008}}, \bibinfo{pages}{004}
  (\bibinfo{year}{2010}), \eprint{0912.1090}.

\bibitem[{\citenamefont{Carlip et~al.}(2009)\citenamefont{Carlip, Deser,
  Waldron, and Wise}}]{Carlip:2008jk}
\bibinfo{author}{\bibfnamefont{S.}~\bibnamefont{Carlip}},
  \bibinfo{author}{\bibfnamefont{S.}~\bibnamefont{Deser}},
  \bibinfo{author}{\bibfnamefont{A.}~\bibnamefont{Waldron}}, \bibnamefont{and}
  \bibinfo{author}{\bibfnamefont{D.~K.} \bibnamefont{Wise}},
  \bibinfo{journal}{Class. Quant. Grav.} \textbf{\bibinfo{volume}{26}},
  \bibinfo{pages}{075008} (\bibinfo{year}{2009}), \eprint{0803.3998};
  \bibinfo{journal}{Phys. Lett.} \textbf{\bibinfo{volume}{B666}},
  \bibinfo{pages}{272} (\bibinfo{year}{2008}), \eprint{0807.0486}.

\bibitem[{\citenamefont{Grumiller and Johansson}(2008)}]{Grumiller:2008qz}
\bibinfo{author}{\bibfnamefont{D.}~\bibnamefont{Grumiller}} \bibnamefont{and}
  \bibinfo{author}{\bibfnamefont{N.}~\bibnamefont{Johansson}},
  \bibinfo{journal}{JHEP} \textbf{\bibinfo{volume}{07}}, \bibinfo{pages}{134}
  (\bibinfo{year}{2008}), \eprint{0805.2610}.

\bibitem[{\citenamefont{Ba\~nados et~al.}(1992)\citenamefont{Ba\~nados,
  Teitelboim, and Zanelli}}]{Banados:1992wn}
\bibinfo{author}{\bibfnamefont{M.}~\bibnamefont{Ba\~nados}},
  \bibinfo{author}{\bibfnamefont{C.}~\bibnamefont{Teitelboim}},
  \bibnamefont{and} \bibinfo{author}{\bibfnamefont{J.}~\bibnamefont{Zanelli}},
  \bibinfo{journal}{Phys. Rev. Lett.} \textbf{\bibinfo{volume}{69}},
  \bibinfo{pages}{1849} (\bibinfo{year}{1992}), \eprint{hep-th/9204099}.
  
  \bibitem[{\citenamefont{Bagchi et~al.}()\citenamefont{Bagchi, Detournay, Fareghbal and Simon}}]{BDRS}
\bibinfo{author}{\bibfnamefont{A.}~\bibnamefont{Bagchi}},
  \bibinfo{author}{\bibfnamefont{R.}~\bibnamefont{Fareghbal}},
  \bibinfo{author}{\bibfnamefont{S.}~\bibnamefont{Detournay}},
  \bibnamefont{and}
  \bibinfo{author}{\bibfnamefont{J.}~\bibnamefont{Simon}},
  \eprint{1208.4372}.

  
  \bibitem[{\citenamefont{Barnich}()\citenamefont{Barnich}}]{BarnichToapp}
\bibinfo{author}{\bibfnamefont{G.}~\bibnamefont{Barnich}},
  \eprint{1208.4371}.

\bibitem[{\citenamefont{Maloney and Witten}(2010)}]{Maloney:2007ud}
\bibinfo{author}{\bibfnamefont{A.}~\bibnamefont{Maloney}} \bibnamefont{and}
  \bibinfo{author}{\bibfnamefont{E.}~\bibnamefont{Witten}},
  \bibinfo{journal}{JHEP} \textbf{\bibinfo{volume}{1002}}, \bibinfo{pages}{029}
  (\bibinfo{year}{2010}), \eprint{0712.0155};
\bibinfo{author}{\bibfnamefont{A.}~\bibnamefont{Maloney}},
  \bibinfo{author}{\bibfnamefont{W.}~\bibnamefont{Song}}, \bibnamefont{and}
  \bibinfo{author}{\bibfnamefont{A.}~\bibnamefont{Strominger}},
  \bibinfo{journal}{Phys. Rev.} \textbf{\bibinfo{volume}{D81}},
  \bibinfo{pages}{064007} (\bibinfo{year}{2010}), \eprint{0903.4573}.

\bibitem[{\citenamefont{Witten}(2007)}]{Witten:2007kt}
\bibinfo{author}{\bibfnamefont{E.}~\bibnamefont{Witten}}
  (\bibinfo{year}{2007}), \eprint{0706.3359}.

\bibitem[{\citenamefont{Frenkel et~al.}(1984)\citenamefont{Frenkel, Lepowsky,
  and Meurman}}]{Frenkel}
\bibinfo{author}{\bibfnamefont{I.~B.} \bibnamefont{Frenkel}},
  \bibinfo{author}{\bibfnamefont{J.}~\bibnamefont{Lepowsky}}, \bibnamefont{and}
  \bibinfo{author}{\bibfnamefont{A.}~\bibnamefont{Meurman}},
  \bibinfo{journal}{Proc. Natl. Acad. Sci. USA} \textbf{\bibinfo{volume}{81}},
  \bibinfo{pages}{3256} (\bibinfo{year}{1984}).

\bibitem[{\citenamefont{Percacci and Sezgin}(2010)}]{Percacci:2010yk}
\bibinfo{author}{\bibfnamefont{R.}~\bibnamefont{Percacci}} \bibnamefont{and}
  \bibinfo{author}{\bibfnamefont{E.}~\bibnamefont{Sezgin}},
  \bibinfo{journal}{Class.Quant.Grav.} \textbf{\bibinfo{volume}{27}},
  \bibinfo{pages}{155009} (\bibinfo{year}{2010}), \eprint{1002.2640}.

\bibitem[{\citenamefont{Bagchi et~al.}()\citenamefont{Bagchi, Detournay, and
  Grumiller}}]{BDG}
\bibinfo{author}{\bibfnamefont{A.}~\bibnamefont{Bagchi}},
  \bibinfo{author}{\bibfnamefont{S.}~\bibnamefont{Detournay}},
  \bibnamefont{and}
  \bibinfo{author}{\bibfnamefont{D.}~\bibnamefont{Grumiller}},
  \bibinfo{note}{work in progress}.




\end{thebibliography}

\end{document}